\documentstyle[12pt,psfig]{article}
\newcommand{\beq}{\begin{equation}}
\newcommand{\eeq}{\end{equation}}
\newcommand{\beqa}{\begin{eqnarray}}
\newcommand{\eeqa}{\end{eqnarray}}
\def\nue{{$\nu_e~$}}

\def\numu{{$\nu_{\mu}$}}

\def\nutau{{$\nu_{\tau}$}}

\def\br{{$^{8}{B} ~$}}

\newcommand{\dm}{\mbox{$\Delta{m}^{2}$~}}

\begin{document}
\begin{flushright}
June 2001\\
\end{flushright}
\vskip 5pt

\begin{center}
{\large {\bf Impact of the  first SNO results on Neutrino Mass and
Mixing }} \vskip 10pt Abhijit Bandyopadhyay\footnote{
abhi@theory.saha.ernet.in}, Sandhya
Choubey\footnote{sandhya@theory.saha.ernet.in}, Srubabati
Goswami\footnote{sruba@theory.saha.ernet.in, currently at Physical
Research Laboratory, Ahmedabad, India }, Kamales Kar\footnote{
kamales@tnp.saha.ernet.in}
\\ \vskip 6pt
 Saha
Institute of Nuclear Physics,\\1/AF, Bidhannagar, Calcutta 700
064, INDIA.\\

{\bf ABSTRACT}

\end{center}


We investigate the implications of the SNO charged-current (CC)
and electron scattering (ES) measurements of  solar \br neutrino
fluxes  for neutrino mass and mixing parameters by performing a
global and unified $\chi^2$ analysis of the solar neutrino data in
the framework of two neutrino mixing.  We consider both $\nu_e
-\nu_{active}$ and $\nu_e -\nu_{sterile}$ solutions and perform
(i) analysis of the total rates data of Cl, Ga, SK and SNO
experiments and (ii) global analysis including the total rates
data, the recoil electron spectrum  data of SK and the CC spectrum
observed at SNO. For the $\nu_e-\nu_{active}$ case the inclusion
of the SNO results in the analysis of the total rates
reduces(enhances) the goodness-of-fit (GOF) of the SMA(LMA)
solution. The flat spectrum observed at SK further favours the LMA
solution over the SMA solution and no allowed area is obtained in
the SMA region at 3$\sigma$ level from the global analysis. 
For the $\nu_e -\nu_{sterile}$ case, with the inclusion of the SNO 
results, all the solutions are 
disfavoured with a probability of more than 99\%  from the
total rates analysis while for the global analysis the GOF of these 
become much worse.

\newpage
The Sudbury Neutrino Observatory (SNO) has declared its first
results \cite{sno} on the measurement of solar \br neutrinos
through the CC detection process
\begin{eqnarray}
\nu_e + d \rightarrow p + p +e^-
\label{nued}
\end{eqnarray}
in the heavy water($D_2O$) of SNO. This  reaction is sensitive to
only \nue and the observed $\nu_e$ flux is
\begin{center}
$\Phi_{CC}^{SNO} = 1.75 \pm 0.07(stat) ^{+0.12}
_{-0.11}(sys) \times 10^6 cm^{-2} s^{-1}$
\end{center}
whereas the expectation from  the standard solar model (SSM) of
\cite{bp00} is $5.05 \times 10^{6} cm^{-2} s^{-1}$. SNO also gives
the $^{8}{B}$ flux measured by the electron scattering (ES)
reaction
\begin{eqnarray}
\nu + e \rightarrow \nu + e \label{nuee}
\end{eqnarray}
as
\begin{center}
$\Phi_{ES}^{SNO}
= 2.39 \pm 0.34 (stat)^{+0.16}_{-0.14}(sys) \times 10^6 cm^{-2} s^{-1}$
\label{phiessno}
\end{center}
The reaction (\ref{nuee}) is  sensitive to both $\nu_e$ and
$\nu_{\mu}$ or $\nu_{\tau}$ and the measured flux is in agreement
with that observed by  the SuperKamiokande (SK) detector \cite{sk,sk1258}
via the same reaction
\begin{center}
$\Phi_{ES}^{SK} = 2.32 \pm  0.03 (stat) ^{+0.08}_{-0.07} \times 10^6 cm^{-2}
s^{-1}$ \label{phiessk}
\end{center}
These new generation high statistics experiments thus confirm
the solar neutrino deficit observed in the pioneering Cl
experiment \cite{cl} and subsequently in Kamiokande \cite{kam} and
the low threshold Ga experiments SAGE, GALLEX and GNO \cite{ga}. A
comparison of the \br \nue flux measured by the CC reaction
(\ref{nued}) with the flux of \br neutrinos measured at SK
signifies the presence of a \numu and/or \nutau component in the
solar neutrino flux at 3.3$\sigma$ level. The total \br neutrino
flux derived from a comparison of $\Phi_{CC}^{SNO}$ and the SK
observed flux $\Phi_{ES}^{SK}$ is found to be $5.44\pm0.99 \times
10^6 cm^{-2}s^{-1}$ which is in excellent agreement with the SSM
predictions \cite{bp00}.

In Table 1 we show the latest results for the total rates measured
in Cl \cite{cl}, Ga \cite{ga}, SK (1258 days) \cite{sk1258} and
SNO (CC and ES) experiments with respect to (w.r.t) the SSM fluxes
of BPB00 \cite{bp00}. The numbers in the parentheses for SK and
SNO (ES) are when the \numu or \nutau  contributions are subtracted. We
also show the composition of the major fluxes in each of these
experiments. 
For the Ga rates we give the weighted average of
SAGE, GALLEX and GNO. 
Apart from the total rates SNO also gives the
CC spectrum of the \br neutrinos and they do not report any
significant distortion with energy. SK has published the data
on the recoil electron energy spectrum in 
separate day and night bins and also
the zenith angle distribution of events 
\cite{skspec,skzenith}. They do not
find any significant variation of the data with energy and
although there is a slight excess of the number of events observed
in the night-time when the neutrinos are passing through the
earth's matter, the effect is only at 1.3$\sigma$.

Various particle physics solutions assuming non standard neutrino
properties have been considered to account for the deficit
\cite{other,mswdk}. The simplest possibility is two flavor
neutrino oscillation which requires \nue to mix with some other
flavor of neutrino. But even in this scenario there are several
disconnected allowed regions in the mass-squared difference -
mixing angle parameter space consistent with the global solar
neutrino data. The flat recoil electron energy spectrum observed
at SK has been responsible in creating a vast change in the
allowed oscillation regions and their goodness of fit (GOF) as
compared to those obtained from analysis of total rates only
\cite{bks98} -\cite{chitre}. The best-fit to the data on total
rates in Cl, Ga and Kamiokande and/or SuperKamiokande experiments
was coming in the MSW \cite{msw} Small-Mixing-Angle (SMA) region.
But with the flat electron energy spectrum observed in SK the
best-fit in the global analysis of rates and spectrum data shifted
to the Large-Mixing-Angle (LMA) region. The fit in the LOW region
(low $\dm \sim$ $10^{-7}$ eV$^2 - 10^{-8}$ eV$^2$ ), where earth
matter effect regenerates the low energy neutrinos also became
good. From the total rates data vacuum oscillation (VO) of
neutrinos were allowed with best-fit \dm $\sim 8.5 \times 10^{-11}$ eV$^2$. 
But in the global analysis with the SK electron
spectrum data
 this became largely
disfavoured as the energy dependence of the survival probability
in this region picked up conflict with the flat electron recoil
energy spectrum.  Recent analysis by SK \cite{sk2001} and other
groups \cite{bks2001,chitre} do find good fits in vacuum
oscillation region for $\dm$ $\sim 4-5 \times 10^{-10}$ eV$^2$
where the energy averaging over the bins smears out the energy
dependence of the probability and the flat spectrum observed in SK
can be accounted for. However the allowed regions are very tiny
around the \dm values in the vacuum oscillation region as well as
somewhat fragile depending on the method of data fitting followed
\cite{bks2001} unlike the MSW allowed regions which are quite
robust against these changes. Apart from these pure MSW and pure
vacuum regions, a grey zone ($\dm \sim 5 \times 10^{-10} $eV$^2$ -
$10^{-9}$ eV$^2$) called the Quasi-Vacuum-Oscillation (QVO) regime
is allowed. For this area of the parameter space
both matter effects inside the sun and the effects due to coherent
oscillation phases are important. Thus there is a continuity in
the allowed parameter regions and the older practice of separate
analysis of the data in vacuum and MSW regions were replaced by
what is called unified analysis which uses a
general expression for probability valid in the whole mass range
$10^{-12} - 10^{-3}$ eV$^2$.
The cutoff in
the $\dm$ from above is due to the constraint from the CHOOZ
reactor experiment \cite{chooz}. Another new aspect was the
appearance of the dark zones ($\theta > \pi/4)$ \cite{murayama}.
In the background of this picture emerging out from detailed
analysis of the available solar neutrino data several studies had
been made on the expectations and implications of the SNO results
\cite{bks2001},\cite{bkssno1}
 -\cite{bargersno}. Now work has started to find the allowed values of mass
squared differences and mixing parameters by actually
incorporating the SNO results in the oscillation analysis \cite
{flsno,bcc,strumia}.

In this paper we investigate the  significance of the SNO results
for neutrino mass and mixing parameters by including these in the
$\chi^2$-analysis of the global solar neutrino data on total rates
in Cl, Ga and SK experiments and the SK day-night recoil electron
spectrum. The definition of $\chi ^2$ used by us is,
\begin{equation}
\chi^2 =
\sum_{i,j} \left(F_i^{th} -
F_i^{exp}\right)
(\sigma_{ij}^{-2}) \left(F_j^{th} - F_j^{exp}\right)
\label{ratefit}
\end{equation}
where i,j runs over the experimental data points. Here
$F_{i}^{\alpha}= \frac{T_i^{\alpha}}{T_{i}^{BPB OO}}$ where
$\alpha$ is $th$ (for the theoretical prediction) or $exp$ (for
the experimental value) and $T_i$ is the total rate in the $i$th
experiment. We first do an analysis with the total rates given in
Table 1. 
The error matrix $\sigma_{ij}$ contains the
experimental errors, the theoretical errors and their
correlations.  For evaluating the error matrix for the total rates
case we use the procedure described in \cite{flap}. The details of
the code used by us can be found in \cite{mswdk,gmr1,gmr2}. For
the rate of $\nu_e-d$ CC events recorded in the SNO detector we
use \beqa R_{CC} = \frac {\int dE_\nu \lambda_{\nu_e}(E_\nu)
\sigma_{CC}(E_\nu)\langle P_{ee}\rangle} {\int dE_\nu
\lambda_{\nu_e} (E_\nu) \sigma_{CC}(E_\nu)} \eeqa \beqa
\sigma_{CC} = \int_{E_{A_{th}}}dE_A\int_0^\infty dE_TR(E_A,E_T)
\frac{d\sigma_{\nu_e d} (E_T,E_\nu)}{dE_T} \label{ccosc}
 \eeqa
where $\lambda_{\nu_e}$ is the normalized $^8B$ neutrino spectrum,
$\langle P_{ee}\rangle$ is the time averaged $\nu_e$ survival
probability, d$\sigma_{\nu_e d}$/dE$_T$ is the differential cross
section of the $\nu_e-d$ interaction, $E_T$ is the true and
$E_A$ the apparent(measured) total energy of the recoil
electrons, $E_{A_{th}}$ is the detector threshold energy which we
take as (6.75+$m_e$) MeV, where $m_e$ is the rest mass of the electron
and R($E_A$,$E_T$) is the energy resolution
function for which we use the expression in \cite{sno}. One of the
major uncertainties in the SNO CC measurement stems from the
uncertainty in the $\nu_e -d$ cross-section. We use the
cross-sections from \cite{nakamura} which are in agreement with
\cite{butler}. Both calculations give an uncertainty of 3\% which
is also the value quoted in \cite{sno}\footnote {It was recently
pointed out in \cite{beacom} that the calculation of both
\cite{nakamura} and \cite{butler} underestimate the total $\nu_e
-d$ cross-section by 6\%. We have not included this effect in our
calculation.}.

The expression for $\nu_e$ survival probability according to an
unified formalism over the mass range $10^{-12} - 10^{-3}$ eV$^2$
and for the mixing angle $\theta$ in the range [0,$\pi/2$] is well
documented \cite{petcov,lisi,murayama} and can be expressed as
\beqa P_{ee}&=&P_{\odot}P_{\oplus} + (1-P_{\odot}) (1-P_{\oplus})
\nonumber \\ && + 2\sqrt{P_{\odot}(1-P_{\odot})
P_{\oplus}(1-P_{\oplus})}\cos\xi \label{probtot} \eeqa where
$P_{\odot}$ denotes the probability of conversion of $\nu_e$ to
one of the mass eigenstates in the sun and $P_{\oplus}$ gives the
conversion probability of the mass eigenstate back to the $\nu_e$
state in the earth. All the phases involved in the Sun, vacuum and
inside Earth are included in $\xi$. This most general expression
reduces to the well known MSW  (the phase $\xi$ is large and
averages out) and vacuum oscillation limit (matter effects are
absent and the phase $\xi$ is important)  for appropriate values
of $\Delta m^2/E$. The procedure which we use for calculating
$P_{\oplus}$ and $P_{\odot}$ in MSW, vacuum as well as the 
in-between quasi-vacuum (QVO) regions  where both $\xi$ and matter
effects are relevant is discussed in \cite{chitre}.

The results for the analysis of total rates for $\nu_e-
\nu_{active}$ oscillations are presented in Table 2 for both pre-SNO 
and post-SNO cases. As far as the pre-SNO
total rates are concerned both SMA and vacuum oscillation give
good fits with the best-fit coming in the SMA region. For post-SNO
the best-fit comes in the VO region. However the noticeable thing
is that with the inclusion of the SNO data the GOF of both SMA and
VO becomes worse and that in the LMA region becomes better.
Prior to SNO, at the best-fit point obtained in the SMA region,
the observed Cl and Ga rates were described very well but the
predicted SK rate was higher. With the introduction of SNO CC rate, the 
best-fit shifts towards higher \dm and higher $\tan^2\theta$, which
corresponds to a lower survival probability for the \br neutrinos
thus lowering the SK and SNO rate. But this also lowers the Cl
rate and the over all $\chi^2$ becomes high.

In the LMA region the survival probabilities of the high energy
neutrinos are given as \cite{concha} \beq P_{ee}^{LMA} \approx
\frac{1}{2}(1 - \epsilon) + f_{reg} \label{peelma} \eeq where
$\epsilon = \cos 2\theta$ and $f_{reg}=P_{2e} - \sin^2 \theta$,
$P_{2e}$ being the probability of $\nu_2 \rightarrow \nu_e$ 
conversion inside the Earth. 
Since the observations of three of the experiments 
(Cl, SK and SNO) which are mainly
sensitive to \br neutrinos are now close, they can be well
described through a single eq. (\ref{peelma}) and the GOF of the
LMA solution becomes better. For low energies relevant for Ga the
matter effects are weak and \beq P_{ee}^{LMA} \approx
\frac{1}{2}(1 + \epsilon^2) \label{peelow} \eeq which gives a
greater probability as compared to eq.(\ref{peelma}) for the same
$\epsilon$ and the Ga rate of Table 1 is accounted for. There is
no significant improvement for the LOW solution for which the
probability is given by eq. (\ref{peelow}) for all energies. In
Table 2 we also give the GOF of the Just So$^{2}$ solution
\cite{justso}. In this region  one gets a very small survival
probability for the $^{7}{Be}$ neutrinos while for the \br
neutrinos the survival probability is close to 1.0 \cite{justso2}.
Since this scenario does not  give any suppression of the \br flux
it gets disfavoured with a probability of more than 99\% by our
total rates analysis with the \br flux 
normalization fixed at the BPB00 SSM value. We have also displayed in 
Table 2 the results of the $\chi^2$ analysis including the SNO ES rate 
in addition to the SNO CC rate. 
The inclusion of the SNO ES rate in the
analysis improves the overall quality of the fits for all the
solutions but it still has large statistical error
and does not make any significant difference between the relative
fit of various solutions.

In Figs. 1 and 2 we plot the
allowed regions for pre-SNO and post-SNO (excluding ES scattering)
respectively at 90\% ($\chi^2 \leq \chi^2_{min} +4.61$), 95\%
($\chi^2 \leq \chi^2_{min} +5.99$), 99\% ($\chi^2 \leq
\chi^2_{min} +9.21$) and 99.73\% C.L.($\chi^2 \leq
\chi^2_{min}+11.83)$ from an analysis of total rates. Since the
GOF of the SMA solution becomes worse with the inclusion of the
SNO CC rate the SMA region reduces in size in Fig. 2. Also it
shifts towards higher values of $\tan^2 \theta$. On the other hand
the allowed  area in the LMA region becomes slightly bigger in the
post-SNO case as the GOF in the LMA region improves. In the LOW
region we get allowed areas at 95\% C.L. for the post-SNO case.

In Table 3  we present the best-fit values of parameters,
$\chi^2_{min}$ and the GOF of the solutions for the
$\nu_e-\nu_{sterile}$ solution from an analysis of total rates.
The GOF in the SMA region goes down from
16.04\% (pre-SNO) to 0.03\% after including the SNO CC rate. Since
the observed SNO CC rate is significantly lower than 
the observed ES rate at SK, pure
$\nu_e -\nu_{sterile}$ transitions are highly disfavoured and this
is responsible for the bad fit obtained in Table 3 after including
the SNO results.

For the global analysis the total $\chi^2$ is defined as \beq
\chi^2 = \chi^2_{rates} + \chi^2_{skspec} + \chi^2_{snospec}
\eeq where $\chi^2_{skspec}$ and $\chi^2_{snospec}$ are the
$\chi^2$ for the SK recoil electron spectrum and SNO CC spectrum
respectively and $\chi^2_{rates}$ corresponds to the $\chi^2$ from
the total rates data. For the calculation of the rates part i,j runs from
1 to 4 if we do not include the ES rate measured in SNO and 1 to 5
if we include the ES rate from SNO; for the SK spectrum part i,j
runs from 1 to 38 corresponding to 19 day and 19 night bins; for
the SNO CC spectrum i,j runs from 1 to 11. To account for the fact
the  ES rate measured in SK  is not independent of the spectrum we
vary the normalization of the spectrum as a free parameter.
Similarly for SNO CC spectrum we introduce a free normalization
to avoid overcounting with the total CC rate.
For the calculation of the error matrix for the SK spectrum
we include the
statistical error, correlated and uncorrelated systematic errors
and the error due to the calculation of the spectrum
\cite{skspec,sk2001}. For the SNO CC spectrum we include the
statistical error and the correlated systematic errors from
\cite{sno}. For all our analyses presented in this paper we
keep the \br flux normalization fixed at SSM value.

The no-oscillation $\chi^2/d.o.f$ is 100.31/52 which is disfavored
at 99.99\% C.L. from the global data. In Table 4 we show the
results of global analysis of the rates and the spectrum data for
oscillation to an active flavour.
To highlight the impact of the
SNO data we present the results for cases with and without SNO\footnote{
The pre-SNO best-fit values for the global analysis are from \cite{chitre}.
The corresponding C.L. contours also appear in \cite{chitre}.}.
For both pre-SNO and
post-SNO we give the best-fit points and the local $\chi^2_{min}$
in five regions -- SMA, LMA, LOW-QVO, VO and Just So$^2$. 

The pre-SNO analysis indicates that
with the inclusion of the SK
day-night spectrum data the GOF of the SMA solution becomes worse
and fit in the LMA and LOW regions become much better, with LMA
giving the best-fit. This worsening of fit in the SMA region is
owing to the fact that the peculiar energy dependence of the
observed rates in Cl, Ga and SK experiments favour larger values
of $\tan^2\theta$
while the flat recoil
electron energy spectrum observed by SK prefers smaller values of
$\tan^2\theta$. LMA and LOW solutions on the other hand can describe the
flat recoil electron spectrum at SK very well and the GOF in these
regions are much better. For the VO case, with the inclusion of the SK
spectrum data the best-fit shifts to $\dm \sim
4.55 \times 10^{-10} eV^2$ for which energy averaging gives an
approximately constant probability for the high energy neutrinos. 
The Just So$^2$ solution, 
although disfavoured from the rates analysis
at more 99\% C.L.,
can explain the flat SK spectrum
well and thus gets allowed from the global analysis at 17.14\%.

With the inclusion of the SNO CC rate into the global analysis, the 
data on total rates demand still 
higher values of $\tan^2\theta$ for the SMA 
solution, thus enhancing the conflict between
the rates and SK spectrum data and the GOF
becomes worse in the SMA 
region\footnote{The contribution from $\chi^2_{rates}$ to
the total $\chi^2$ increases from 6.39 at the SMA best-fit for the
pre-sno case to 14.99 with the inclusion of the SNO CC rate
reducing the overall GOF.}. If we look at the post-SNO
$\chi^2_{min}$ in Table 4 for the case excluding the SNO CC
spectrum and the ES data then we find that the SMA 
solution becomes more disfavoured with SNO, while
LMA, LOW and VO are seen to improve, with the best-fit still 
in the LMA region.
The Just So$^2$ 
solution gets worse with the introduction of the SNO CC 
rate, however it is still allowed with a probability of 8.1\%. 

We have repeated the post-SNO 
global analysis by including the SNO ES rate and 
the SNO CC spectrum in addition to the SNO CC rate and have presented the 
results in Table 4. We find that due to large errors, both 
statistical and systematic, the effect of addition of the SNO CC
spectrum in the analysis is to increase the $\chi^2/d.o.f$ and
hence reduce the GOF for all the solutions in general.



In Fig. 3 we show the allowed regions at 90\%, 95\%, 99\%  and
99.73\% C.L. obtained from the global analysis for
$\nu_e-\nu_{active}$ transitions including {\it all published SNO
data.} The significant change in the allowed regions after
including the SNO results is the disappearance of the SMA region
even at 99.73\% C.L. ($3\sigma$) 
as a result of increased conflict between the
total rates and SK spectrum data.
For the Just So$^2$ solution also there is no
allowed region at 99.73\% C.L. after including the SNO data in
the global analysis of rates and SK spectrum{\footnote
{From Table 4 we see that for the pre-SNO case we have allowed 
area at 99.73\% C.L. in the Just So$^2$ region.}}.

In Table 5  we present the results of global analysis for $\nu_e -
\nu_{sterile}$ solution  and as expected  the fits become worse
with the inclusion of SNO results. The SMA and the VO solutions
which were allowed at 22.9\% and 32.57\% respectively without the
SNO results are now allowed at only 5.12\% and 6.10\%. The GOF in
the LMA and LOW regions also become worse. Prior to the SNO
results the SMA and VO was giving much better fit to the global
data as compared to the LMA and LOW solutions since the former
could account for the total rates data much better. But with the
inclusion of the SNO CC rate the GOF of the SMA and VO solutions
for the total rates analysis is reduced by a large amount and as
a result all the solutions become more disfavoured for the sterile
neutrino case.


The GOF of the SMA solution 
is very sensitive to the uncertainty of the $\nu_e-d$
cross-sections used. To illustrate this point in Table 6 we give
the GOF of the various solutions for the $\nu_e -\nu_{active}$
case using the $\nu_e -d$ cross-sections from \cite{lisibahcall}
and a conservative estimate of uncertainty of 9\%. Comparing the
GOF in the LMA and SMA region from the global 
analysis we find that with the use of a 9\%
uncertainty in the $\nu_e-d$ cross-sections the SMA region gets
allowed at the $3\sigma$ level.

To summarise, we include the recent SNO results in global $\chi^2$
analysis of the solar neutrino data assuming $\nu_e$ to mix with
either another active neutrino or a sterile neutrino. We first
perform a fit to the total rates including (i) the SNO CC rate and
(ii) both SNO CC and SNO ES rates, along with the total rates from
Cl, Ga and SK experiments. For the $\nu_e-\nu_{active}$ case, 
SMA, LMA, LOW and VO solutions which
were allowed from pre-SNO total rates analysis are still allowed
but the inclusion of the SNO CC data in the analysis of {\it total
rates} worsens the GOF of the SMA and VO solution and betters the
GOF of the LMA solution. 
The inclusion of the SNO CC rate
disfavours all the solutions for pure $\nu_e-\nu_{sterile}$ case
with a probability of more than 99\%.

We next perform  a global analysis  of rates and spectrum data
including
(i)the SNO CC rate and (ii) the SNO CC and ES rates and the SNO CC spectrum
along with
the total rates of Cl, Ga and SK experiments and the SK day-night spectrum.
 For $\nu_e
-\nu_{active}$
 case, the global analysis gives five allowed
solutions -- LMA, VO, LOW, SMA, Just So$^2$ -- in order of decreasing
GOF. However no allowed area is obtained at 99.73\% C.L. for SMA
and Just So$^2$ solutions. 
With the inclusion of the SNO CC rate the mismatch between the best-fit
parameters for the rates and SK spectrum increases in the SMA region while
the LMA gives a very good fit to the global data. This results in a
marked improvement of the LMA solutions over the SMA solution and we get
no allowed area in the SMA region even at the $3\sigma$ level.

The sterile neutrino alternative gets highly disfavoured by the
rates analysis and the global analysis gives a GOF of only about 5\% in
the SMA region. However for an arbitrary \br flux normalization, a
small admixture with the sterile neutrino state cannot be ruled
out completely as is shown by the model-independent analysis
performed in \cite{bmw}.

{\it Note added:} Our paper (hep-ph/0106264) appeared on the net
at about the same time as \cite{flsno} and \cite{bcc}. We have
updated our calculation with the latest $\nu_e-d$ cross-sections
from \cite{nakamura}. Our method of analysis is same as in
\cite{flsno} but we have included transitions to sterile neutrinos
not included in \cite{flsno}. For the active case our results
agree with \cite{flsno}. 
The analysis in \cite{bcc} uses a somewhat different definition of 
$\chi^2$ and they include an extra parameter to 
determine the active-sterile admixture in their analysis. 
We have also presented results including the SNO ES rate and 
the SNO CC spectrum data in our analysis, not
included in the analyses of \cite{flsno} and \cite{bcc}.

\vskip 10 pt

The authors would like to thank D.P. Roy for his valuable comments
and suggestions. S.G. would like to thank the theory group of
Physical Research Laboratory for their hospitality.

\newpage
\begin{table}
\caption{
 The ratio of the observed solar neutrino rates to the
corresponding BPB00 SSM predictions used in this analysis.
}
\[
\begin{tabular}{ccc} \hline
experiment & $\frac{obsvd}{BPB00}$ & composition \\ \hline Cl &
0.335 $\pm$ 0.029 & $B$ (75\%), $Be$ (15\%)
\\
Ga & 0.584 $\pm$ 0.039 &$pp$ (55\%), $Be$ (25\%), $B$ (10\%)
\\
SK & 0.459 $\pm$ 0.017  & $B$ (100\%)
\\
& (0.351 $\pm$0.017) & \\ SNO(CC) & 0.347 $\pm$ 0.027 & $B$
(100\%)
\\ SNO(ES) & 0.473 $\pm$ 0.074 & $B$ (100\%) \\
&(0.368 $\pm$ 0.074 ) & \\ \hline
\end{tabular}
\]
\end{table}

\begin{table}
\caption
{The best-fit values of the parameters,
$\chi^2_{min}$, and the goodness of fit from
an analysis of the total rates given in Table 1 for
$\nu_e - \nu_{active}$.}

\begin{center}
\begin{tabular}{cccccc}
\hline
&Nature of & $\Delta m^2$ &
$\tan^2\theta$&$\chi^2_{min}$& Goodness\\
&Solution & in eV$^2$&  & & of fit\\
\hline
&SMA & $5.96 \times 10^{-6}$&$1.39 \times 10^{-3}$ & 0.30 & 58.39\%\\
pre-SNO &LMA & $2.40\times 10^{-5}$ & 0.31 & 2.91 & 8.80\%\\
(Cl+Ga+SK)  &LOW-QVO & $ 1.34 \times 10^{-7}$ & 0.64 & 7.49 & 0.62\%\\
&VO& $8.79\times 10^{-11}$& 0.43 & 0.32  & 57.16\%\\
&Just So$^2$& $5.40\times 10^{-12}$ & 1.00 & 12.86 & $3.36\times10^{-2}$\%\\
\hline
 &SMA & $7.71\times 10^{-6}$&$1.44 \times 10^{-3}$ & 5.44 & 6.59\% \\
 post-SNO &LMA & $2.59 \times 10^{-5}$ & 0.34 & 3.40 & 18.27\% \\
 (Cl+Ga+SK &LOW-QVO & $ 1.46 \times 10^{-7}$ & 0.67 & 8.34 & 1.55\%\\
+ SNO CC ) &VO& $7.73\times 10^{-11}$& 0.27 & 2.49 &28.79\%\\
&Just So$^2$& $5.38\times 10^{-12}$ & 1.29 & 19.26 &$6.57\times10^{-3}$\%\\
\hline
post-SNO&SMA & $7.71 \times 10^{-6}$&$1.44 \times 10^{-3}$ & 5.44 & 14.23\%\\
(Cl+Ga+SK&LMA & $2.32 \times 10^{-5}$ & 0.33 & 3.47 & 32.47\%\\
+SNOCC&LOW-QVO & $ 1.14\times 10^{-7}$ & 0.81 & 9.24 & 2.63\%\\
+SNOES)&VO& $7.74\times 10^{-11}$& 0.27 & 2.92 & 40.41\%\\
&Just So$^2$& $5.38\times 10^{-12}$& 1.27 & 19.42 & $2.24\times10^{-2}$\%\\
\hline
\end{tabular}
\end{center}
\end{table}

\begin{table}
\caption {The best-fit values of the parameters, $\chi^2_{min}$,
and the goodness of fit from an analysis of the total rates given
in Table 1 for $\nu_e - \nu_{sterile}$.}
\begin{center}
\begin{tabular}{cccccc}
\hline
&Nature of & $\Delta m^2$ & $\tan^2\theta$&$\chi^2_{min}$& Goodness\\
&Solution & in eV$^2$&  & & of fit\\
\hline
&SMA & $4.43 \times 10^{-6}$&$1.44 \times 10^{-3}$ & 1.97 & 16.04\%\\
pre-SNO &LMA & $6.41 \times 10^{-5}$ & 0.58 &17.45 & 2.94 $\times 10^{-3}$\%\\
(Cl+Ga+SK)   &LOW-QVO & $ 1.49 \times 10^{-7}$ & 0.85 & 18.01 & 2.19 $\times 10^{-3}$\%\\
& VO & $ 8.99\times 10^{-11}$& 0.36 & 2.70 & 10.03\%\\
&Just So$^2$& $ 5.40\times 10^{-12}$& 1.00 & 12.89 & $3.30 \times 10^{-2}$\%\\
\hline
&SMA & $4.18 \times 10^{-6}$&$5.72 \times 10^{-4}$ &
   17.24& $1.80 \times 10^{-2}$
   \%  \\
post-SNO& LMA & $4.98 \times 10^{-5}$ & 0.54 &
   23.96 & 6.27$\times 10^{-4}$\% \\
(Cl+Ga+SK &LOW-QVO & $1.00\times10^{-7}$ & 0.94 & 24.26 & 5.40$\times10^{-4}$\%
   \\
+SNOCC) &VO& $1.07 \times 10^{-10}$& 0.27 &15.71 & $3.88 \times 10^{-2}$\%\\
 &Just So$^2$& $5.37 \times 10^{-12}$& 1.28 &19.40 & $6.13 \times 10^{-3}$\%\\
 \hline
post-SNO   &SMA & $5.20 \times 10^{-6}$&$4.38 \times 10^{-4}$ &
   17.34& $6.02 \times 10^{-2}$
   \%  \\
(Cl+Ga+SK & LMA & $6.61 \times 10^{-5}$ & 0.55 &
   24.42 & 2.04 $\times 10^{-3}$\% \\
 + SNO CC  &LOW-QVO & $ 2.96\times 10^{-8}$ & 0.87 & 22.16 & 6.04$\times 10^{-3}
$\%
   \\
+ SNO ES) &VO& $7.86 \times 10^{-11}$& 0.23 &23.76 & 2.80$\times 10^{-3}$\%\\
&Just So$^2$& $5.37 \times 10^{-12}$& 1.27 &19.56 & 2.09$\times 10^{-2}$\%\\
 \hline
  \end{tabular}
 \end{center}
\end{table}

\begin{table}
\caption {The best-fit values of the parameters, $\chi^2_{min}$,
and the goodness of fit from the global analysis of rate and
spectrum data 
for $\nu_e -
\nu_{active}$.}
 \begin{center}
  \begin{tabular}{cccccc}
   \hline
   &Nature of & $\Delta m^2$ &
   $\tan^2\theta$&$\chi^2_{min}$& Goodness\\
   &Solution & in eV$^2$&  & & of fit\\
   \hline
   &SMA & $5.48 \times 10^{-6}$&$4.88 \times 10^{-4}$ &
   43.59 & 24.57\%  \\
  pre-SNO &LMA & $5.08 \times 10^{-5}$ & 0.35 &
   34.73 & 62.14\% \\
(Cl+Ga+SK  &LOW-QVO & $ 1.55 \times 10^{-7}$ & 0.66 & 38.50 & 44.66\%
   \\
 + SK spec)  &VO& $4.55\times 10^{-10}$& 0.44 & 37.80 & 47.86\%\\
 &Just So$^2$& $5.43 \times 10^{-12}$& 1.00 & 46.13 & 17.14\% \\
   \hline
post-SNO  &SMA & $5.28 \times 10^{-6}$&$3.75 \times 10^{-4}$ &
   51.14 & 9.22\%  \\
(Cl+Ga+SK &LMA & $4.70 \times 10^{-5}$ & 0.38 &
   33.42 & 72.18\% \\
+SNOCC  &LOW-QVO & $ 1.76 \times 10^{-7}$ & 0.67 & 39.00 & 46.99\%
   \\
+SK spec) &VO& $4.64\times 10^{-10}$& 0.57 &38.28 & 50.25\%
\\ 
   &Just So$^2$& $5.37\times 10^{-12}$& 0.77 & 51.90 & 8.10\% \\
 \hline
     &SMA & 5.29 $\times 10^{-6}$&$3.89 \times 10^{-4}$ &
   65.20 & 7.30\%  \\
  post-SNO &LMA & $4.49 \times 10^{-5}$ & 0.38 &
   47.84 & 56.05\% \\
 (all data)  &LOW-QVO & $ 1.70 \times 10^{-7}$ & 0.66 & 53.30 & 34.85\%
   \\
    &VO& $4.53 \times 10^{-10}$& 0.36 & 56.82 & 23.60\% \\
    &Just So$^2$& $5.37 \times 10^{-12}$& 0.78 & 66.29 & 6.12\% \\
   \hline

  \end{tabular}
  \end{center}
\end{table}

\begin{table}
\caption {The best-fit values of the parameters, $\chi^2_{min}$,
and the goodness of fit from the global analysis of rates and
spectrum data 
for $\nu_e - \nu_{sterile}$.}

 \begin{center}
  \begin{tabular}{cccccc}
   \hline
   &Nature of & $\Delta m^2$ &
   $\tan^2\theta$&$\chi^2_{min}$& Goodness\\
   &Solution & in eV$^2$&  & & of fit\\
   \hline
   & SMA & 4.03 $\times 10^{-6}$&$4.9 \times 10^{-4}$ &
   44.11 & 22.90\%\\
  &LMA & $6.09 \times 10^{-5}$ & 0.56 &
   47.15 & 14.67\% \\
   pre-SNO&LOW-QVO & $ 3.08 \times 10^{-8}$ & 0.85 & 47.16 & 14.65\% \\
    &VO& $4.54\times 10^{-10}$& 0.39 & 41.37 & 32.57\%\\
 &Just So$^2$& $5.39\times 10^{-12}$& 1.00 & 45.61 & 18.51\%\\
    \hline
    & SMA & 3.87 $\times 10^{-6}$&$3.69 \times 10^{-4}$ &
   67.37 & 5.12\%  \\
post-SNO   &LMA & $7.66 \times 10^{-5}$ & 0.48 &
   67.37 & 5.12\% \\
(all data) &LOW-QVO & $ 2.89 \times 10^{-8}$ & 1.00 & 68.19 & 4.45\%\\
    &VO& $4.67\times 10^{-10}$& 0.30 & 66.31 & 6.10\%\\
 &Just So$^2$& $5.37\times 10^{-12}$& 0.78 & 66.48 & 5.93\%\\
    \hline
  \end{tabular}
  \end{center}
\end{table}

\begin{table}[h]
\caption {The best-fit values of the parameters, $\chi^2_{min}$,
and the goodness of fit using ${\nu_e- d}$ 
cross section from $\cite{lisibahcall}$.}
 \begin{center}
  \begin{tabular}{cccccc}
   \hline
   &Nature of & $\Delta m^2$ &
   $\tan^2\theta$&$\chi^2_{min}$& Goodness\\
   &Solution & in eV$^2$&  & & of fit\\
   \hline
  &SMA & $6.13\times10^{-6}$&$1.46\times10^{-3}$ &
   2.97 & 22.65\%  \\
post-SNO  &LMA & $2.30 \times 10^{-5}$ & 0.32 &
   3.26 & 19.59\% \\
(Cl+Ga+SK&LOW-QVO & $ 1.40 \times 10^{-7}$ & 0.70 & 7.88 & 1.93\%
   \\
   +SNOCC)&VO & $7.95\times 10^{-11}$& 0.27 & 2.20 &33.29\% \\
   &Just So$^2$ & $5.37\times 10^{-12}$ & 0.92 & 14.25 & 
      $8.05\times 10^{-2}$\% 
\\ 
    \hline
 post-SNO   &SMA & $5.36\times10^{-6}$&$4.15\times10^{-4}$ &
   45.21 & 22.85\%  \\
(Cl+Ga+SK &LMA & $4.45\times10^{-5}$ & 0.36 &
   34.98 & 65.38\% \\
+SNOCC&LOW-QVO & $1.56\times10^{-7}$ & 0.69 & 38.38 & 49.80\%
   \\
+SK spec) &VO& $4.54\times10^{-10}$& 2.35 &37.83 & 52.31\% \\
  & Just So$^2$ & $5.37\times 10^{-12}$& 0.88 & 46.93&17.94\%
\\ \hline
  \end{tabular}
  \end{center}
\end{table}

\begin{figure}[ht]
\vskip -1.9in
\centerline{\psfig{figure=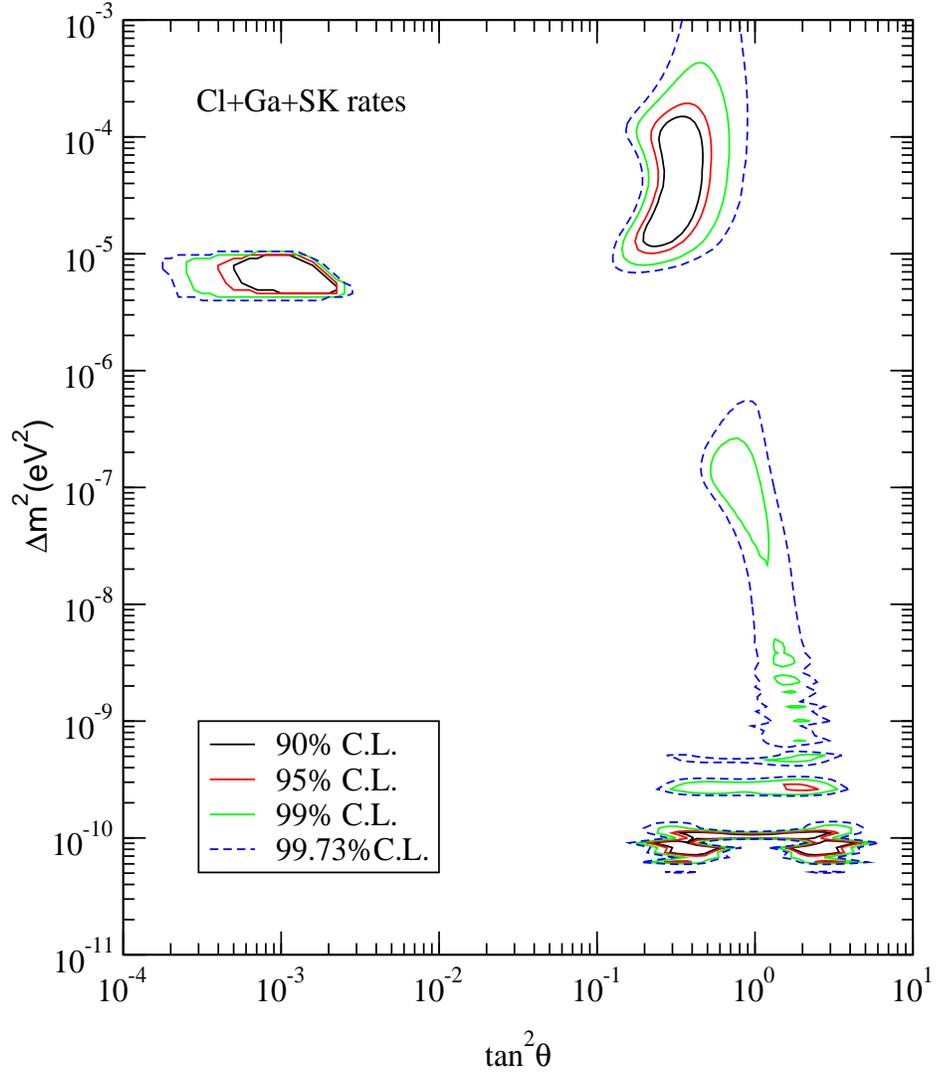,width=15.0cm,height=22.0cm}}
\vskip -1.00in
\caption{
The pre-SNO 90\%, 95\%, 99\%  and 99.73\% C.L. allowed area from the
fit to the data on total rates from the Cl, Ga and SK experiments assuming
two-generation oscillations to active neutrino.}
\end{figure}

~~~~~~~~~~~~~~~~~~
\begin{figure}[ht]
\vskip -1.9in
\centerline{\psfig{figure=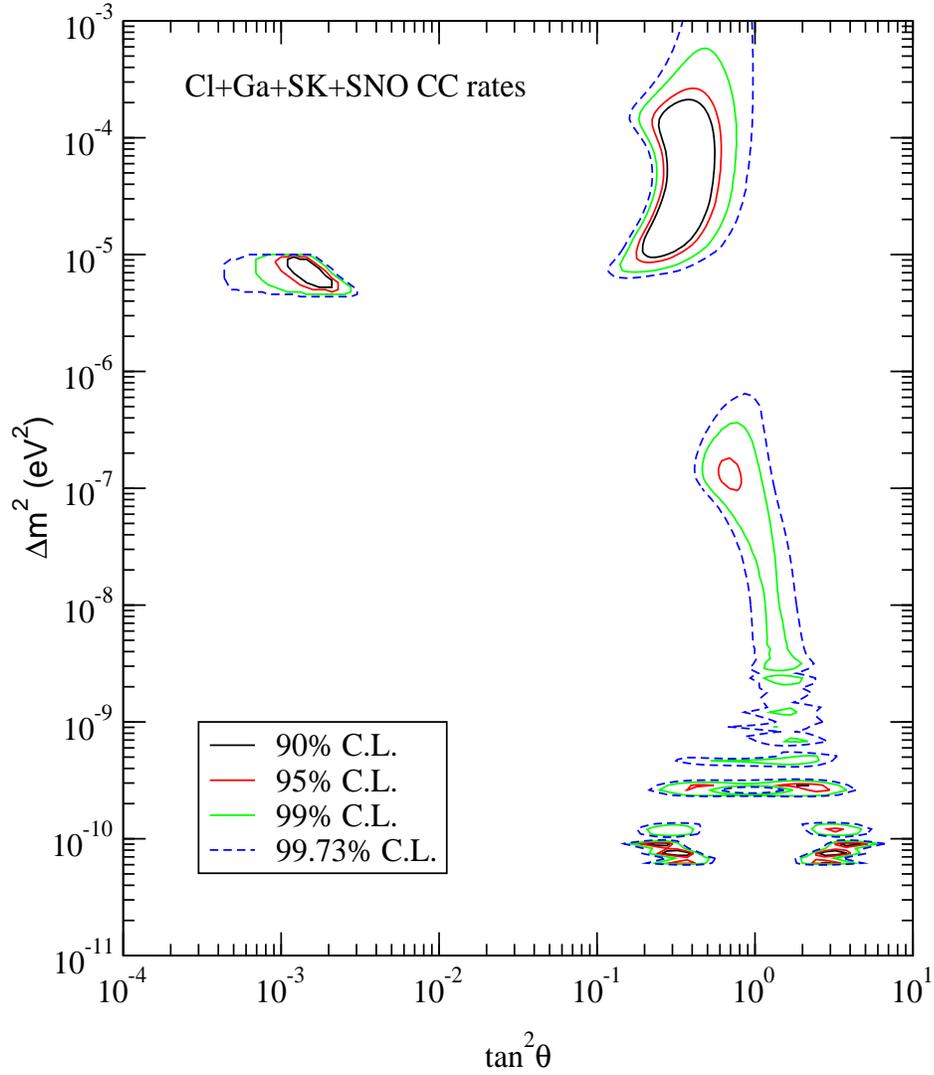,width=15.0cm,height=22.0cm}}
\vskip -1.00in
\caption{
The post-SNO 90\%, 95\%, 99\% and 99.73\% C.L. allowed area from the
fit to the data including the SNO CC rate along with the
total rates from the Cl, Ga, SK experiments for
two-generation oscillations to active neutrino.}
\end{figure}

\begin{figure}
\vskip -1.9in
\centerline{\psfig{figure=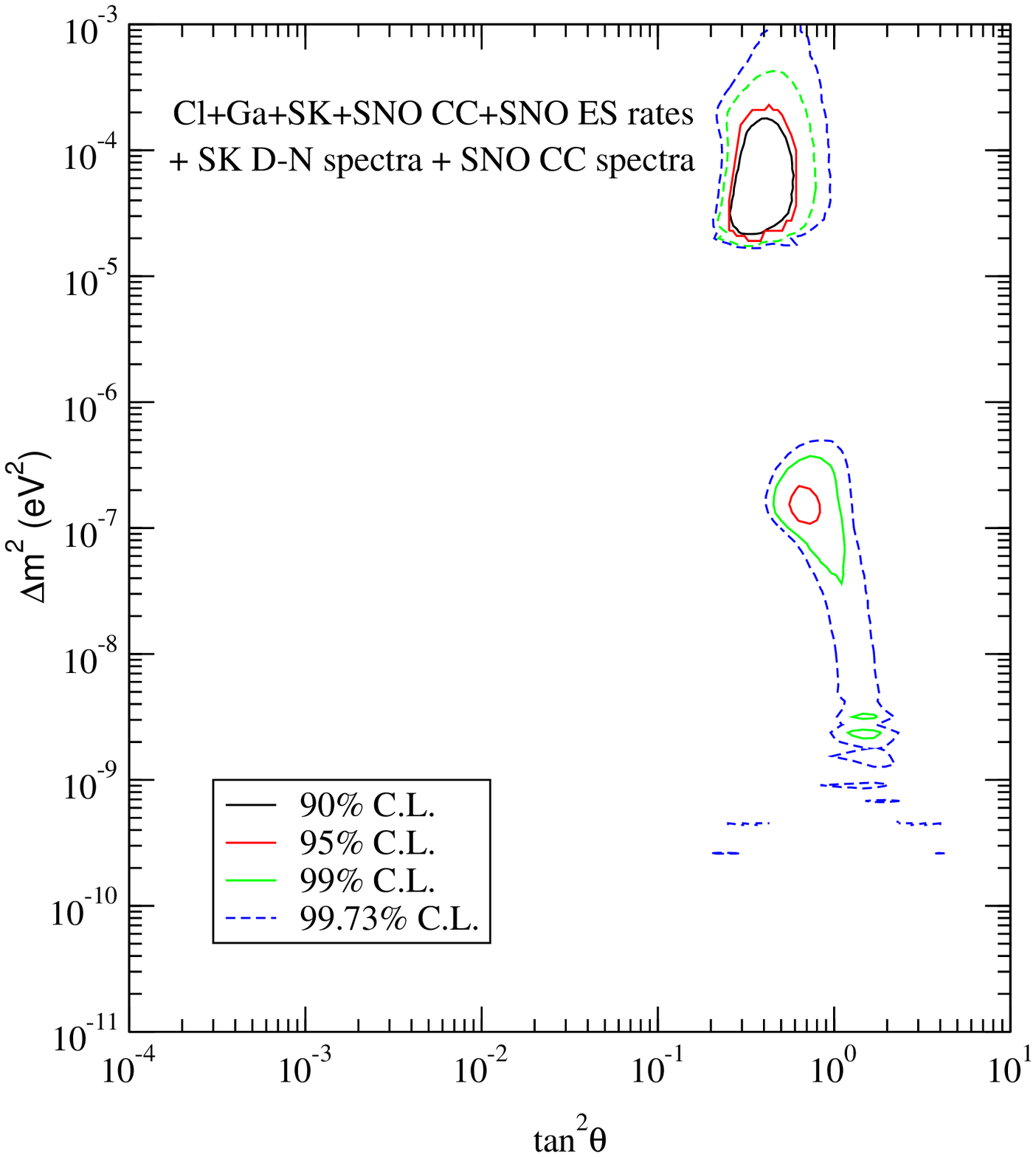,width=15.0cm,height=22.0cm}}
\vskip -1.00in
\caption{
The post-SNO 90\%, 95\%, 99\% 99.73\% C.L. allowed area from the
global analysis of the total rates from Cl, Ga, SK and SNO (both
CC and ES), the 1258 day SK recoil electron energy spectrum
at day and night and the SNO CC spectrum data, assuming
two-generation oscillations to active neutrino.}
\end{figure}


\vskip 3mm
\begin{thebibliography}{99}
\bibitem{sno}
Q.R. Ahmad {\it et al}.,
http://www.sno.phy.quennsu.ca/sno/firstresults/, nucl-ex/0106015.

\bibitem{bp00} J.N. Bahcall, S. Basu, M. Pinsonneault,
Ap. J. {\bf 555}, 990 (2001).

\bibitem{sk}Y. Fukuda {\it et al.} (The Super-Kamiokande collaboration),
Phys. Rev. Lett. {\bf 81}, 1158 (1998); erratum {\bf 81}, 4279
(1998).

\bibitem{sk1258}Y. Fukuda {\it et al.}, Phys. Rev. Lett.
{\bf 86}, 5651 (2001).

\bibitem{cl}B.T. Cleveland {\it et al.} Astrophys. J {\bf 496}, 505 (1998).

\bibitem{kam}Y. Fukuda {\em et al.}, (The
Kamiokande collaboration), {Phys. Rev. Lett.} {\bf 77}, 1683
(1996).

\bibitem{ga}J.N. Abdurashitov {\em et
al.}, (The SAGE collaboration), {Phys. Rev. Lett.} {\bf 77},
4708 (1996); Phys. Rev. {\bf C 60}, 055801 (1999); W. Hampel {\em
et al.}, (The Gallex collaboration), {Phys. Lett.} {\bf B388},
384 (1996); Phys. Lett. {bf B447}, 127 (1999); Talk presented in
Neutrino 2000 held at Sudbury, Canada (T.A. Kirsten for The Gallex
collaboration), Nucl. Phys. {\bf B} Proc. Suppl. {\bf 77}, 26
(2000); M. Altmann {\it et al.}, (The GNO collaboration),Phys.
Lett. {bf B492},16 (2000); Talk presented in Neutrino 2000 held at
Sudbury, Canada ( E. Bellotti for the GNO Collaboration) Nucl.
Phys. {\bf B} Proc. Suppl. {\bf 91} 44 (2001).

\bibitem{skspec}Y. Fukuda {\it et al.} (The Super-Kamiokande collaboration),
Phys. Rev. Lett. {\bf 82}, 2430 (1999).

\bibitem{skzenith}Y. Fukuda {\it et al.} (The Super-Kamiokande collaboration),
Phys. Rev. Lett. {\bf 82}, 1810 (1999).

\bibitem{other}S. Choubey, S. Goswami, D. Majumdar,
Phys. Lett. {\bf B484}, 73 (2000).
O.G. Miranda, C. Pe\~{n}a-Garay, T.I. Rashba, V.B. Semikoz, J.W.F.
Valle, Nucl. Phy. {\bf B595},
360 (2001); D. Majumdar, A. Raychaudhuri and A. Sil,
Phys. Rev. {\bf D63},
073014 (2001); A.M. Gago, H. Nunokawa and R. Zukanovich Funchal,
Nucl. Phys. Proc. Suppl. {\bf 100}, 68 (2001);
M.M. Guzzo, H. Nunokawa, P.C. de Holanda, O.L.G.
Peres, hep-ph/0012089; J. Pulido, hep-ph/0106201.

\bibitem{mswdk}A. Bandyopadhyay, S. Choubey, S. Goswami, Phys. Rev.
{\bf D63}, 113019 (2001).


\bibitem{bks98}J.N. Bahcall, P.I. Krastev and A.Yu. Smirnov,
{Phys. Rev.} {\bf D58}, 096016 (1998).

\bibitem{valle}M.C.Gonzalez-Garcia and C. Pe\~{n}a-Garay,
Nucl. Phys. Proc. Suppl. {\bf 91}, 80 (2000);
 M.C. Gonzalez-Garcia, P.C.  de Holanda, C.
Pe\~{n}a-Garay, and J.W.F. Valle, Nucl. Phys. {\bf B573},
3 (2000).

\bibitem{barger} V. Barger and K. Whisnant, Phys. Lett. {\bf B456},
54 (1999).
\bibitem{gmr1} S. Goswami, D. Majumdar and
A. Raychaudhuri, hep-ph/9909453.

\bibitem{gmr2}S. Goswami, D. Majumdar and
A. Raychaudhuri, Phys. Rev. {\bf D63},
013003 (2001).

\bibitem{bks2001} J.N. Bahcall, P.I. Krastev, and A.Yu. Smirnov,
JHEP {\bf 0105}, 015 (2001).

\bibitem{dp} S. Choubey, S. Goswami, N. Gupta and D.P. Roy, hep-ph/0103318,
to appear in Phys. Rev. {\bf D}.

\bibitem{chitre}S. Choubey, S. Goswami, K. Kar, A.R. Antia, S.M. Chitre,
hep-ph/0106168.

\bibitem{msw}L. Wolfenstein, {Phys. Rev.} {\bf D34}, 969 (1986);
S.P. Mikheyev and A.Yu. Smirnov,  {Sov. J. Nucl. Phys.} {\bf
42(6)}, 913 (1985); {Nuovo Cimento} {\bf 9c}, 17 (1986) .

\bibitem{sk2001}  Y.Fukuda {\it et al.} Phys. Rev. Lett. {\bf 86},
5656 (2001).


\bibitem{chooz}M. Apollonio et al., Phys. Lett. B446, 415 (1999).

\bibitem{murayama} A.de. Gouvea, A. Friedland, H. Murayama,
Phys.Lett.{\bf B490}, 125, (2000).

\bibitem{bkssno1}J. N. Bahcall, P.I. Krastev
, A. Yu. Smirnov
JHEP {\bf 0105}, 015 (2001).

\bibitem{bkssno2}J.N. Bahcall
, P. I. Krastev,
A. Yu. Smirnov
Phys. Rev. {\bf D63}, 053012 (2001).

\bibitem{bkssno3}J. N. Bahcall, P.I. Krastev
, A. Yu. Smirnov, Phys. Rev. {\bf D62}, 093004, (2000).

\bibitem{bargersno}V. Barger, D. Marfatia, K. Whisnant and B.P. Wood,
hep-ph/0104095.

\bibitem{flsno}G.L. Fogli, E. Lisi, D. Montanino and A. Palazzo,
hep-ph/0106247.

\bibitem{bcc}J.N. Bahcall, M.C. Gonzalez-Garcia and C. Penya-Garay,
hep-ph/0106258.

\bibitem{strumia}P. Creminelli, G. Signorelli, A. Strumia, 
hep-ph/0102234 (updated version, july 2000).

\bibitem{flap}G.L. Fogli and E. Lisi, {Astropart. Phys.} {\bf 3},
185 (1995).

\bibitem{nakamura} S. Nakamura, T. Sato, V. Gudkov and
K. Kubodera, Phys. Rev. {\bf C63}, 034617 (2001).

\bibitem{butler}M. Butler, J. Chen and X. Kong, Phys. Rev. {\bf C63},
034617, (2001).

\bibitem{beacom} J.F. Beacom and S.J. Parke, hep-ph/0106128.

\bibitem{petcov}S.T. Petcov, Phys. Lett. {\bf B214}, 139, (1988);
,{\bf 200},373, (1988); S.T. Petcov and J. Rich, Phys. Lett. {\bf B426},
(1989).

\bibitem{lisi} G.L. Fogli, E. Lisi, D. Montanino, A. Palazzo
, Phys. Rev. {\bf D62}, 113004, (2000).

\bibitem{concha}M.C. Gonzalez-Garcia, C. Pe\~{n}a-Garay, Y. Nir and 
A. Yu. Smirnov, Phys. Rev. {\bf D63}, 013007 (2001).  

\bibitem{justso}R.S. Raghavan, Science {\bf 267}, 45 (1995); P.I.
Krastev and S.T. Petcov, Phys. Rev. {\bf D53}, 1665 (1996).

\bibitem{justso2}See Fig. 5 of \cite{bks2001}.

\bibitem{lisibahcall}http://www.sns.ias.edu/$\sim$jnb/, 
J.N. bahcall and E. Lisi Phys. Rev. {\bf D54}, 5417 (1996).

\bibitem{bmw}V. Barger, D. Marfatia and K. Whisnant, hep-ph/0106207.


\end{thebibliography}
\end{document}